\documentclass[12pt]{article}
\title{Correlator of Topological Charge Densities\\
in Instanton Model in QCD.}
\vspace{2cm}
\author{B.L.\,Ioffe\thanks{e-mail:ioffe@vitep5.itep.ru}$\;$ and
A.V.\,Samsonov\thanks{e-mail:sams@heron.itep.ru}\\
\\
Institute of Theoretical and Experimental Physics\\
B.Cheremushkinskaya 25, 117218, Moscow, Russia}
\date{}
\begin{document}

\maketitle

\newcommand{\be}{\begin{equation}}
\newcommand{\ee}{\end{equation}}

\def\la{\mathrel{\mathpalette\fun <}}
\def\ga{\mathrel{\mathpalette\fun >}}
\def\fun#1#2{\lower3.6pt\vbox{\baselineskip0pt\lineskip.9pt
\ialign{$\mathsurround=0pt#1\hfil##\hfil$\crcr#2\crcr\sim\crcr}}}

\begin{abstract}

The QCD sum rule for the correlator of topological charge densities 
$\chi(Q^2)$
and related to it  longitudinal part of the correlator of singlet axial
currents is considered in the framework of instanton model. The coupling
constant $f_{\eta^{\prime}}$  of $\eta^{\prime}$-meson with 
the singlet axial current is
determined. Its value appears to be in a good coincidence with the value
determined recently from the connection of the part  of proton spin
$\Sigma$, carried by $u,d,s$ quarks, with the derivative of QCD topological
susceptibility $\chi^{\prime}(0)$. From the same sum rule
$\eta-\eta^{\prime}$  mixing angle $\theta_8$ is found in the framework of
two mixing angles model. The value of $\theta_8$ is close to that found in
the chiral effective theory. The correlator of topological charge densities
$\chi(Q^2)$  at large $Q^2$  is calculated and it was found that its 
$Q^2$-dependence matches well with $Q^2$-dependence at low $Q^2$
determined by
the known $\chi^{\prime}(0)$ and by contributions of $\pi$- and
$\eta$-mesons.

\end{abstract}

\newpage

\section{Introduction.}

Recently the vacuum expectation value of singlet axial current $f^2_0$
induced by external singlet axial field $A_{\mu}$  has been found [1]:

\be
\langle 0\mid j_{\mu 5}\mid 0\rangle_A = 3f^2_0 A_{\mu}\,.
\label{1}
\ee
In (\ref{1})  $j_{\mu 5}$ is the singlet quark current

\be
j_{\mu 5}(x) =\sum_{q}  \bar{q}(x)\gamma_{\mu}\gamma_5 q(x),~~~~q=u,d,s.
\label{2}
\ee
The term 

\be
\Delta L = j_{\mu 5} A_{\mu}
\label{3}
\ee
was added to QCD Lagrangian, where  $A_{\mu}$ is a constant in space and time singlet
axial field and the limit of weak $A_{\mu}$  field was considered. It was
found in the limit of massless $u,d,s$ quarks \cite{1,2}:

\be
f^2_0 = (2.8\pm 0.7) \cdot 10^{-2} ~GeV^2.
\label{4}
\ee
This result was obtained, constructing the QCD sum rule in the external field
$A_{\mu}$ to determine the part of proton spin $\Sigma$ carried by $u,d,s$
quarks and related to the proton matrix element of the current $j_{\mu 5}$:

\be
2m s_{\mu}\Sigma = \langle p,s \mid j_{\mu 5}\mid p,s \rangle,
\label{5}
\ee
where $s_{\mu}$ is the proton spin 4-vector, $m$  is the proton  mass. The
sum rule for $\Sigma$  essentially depends on $f^2_0$  and the numerical
result (4)  comes in two ways, resulting in the same value:

1)  from the requirement of consistency of phenomenological and calculated
in QCD sides of the sum rule as functions of the Borel parameter $M^2$;

2) by using for $\Sigma$ its experimental value, $\Sigma=0.3\pm 0.1$.

As was shown in \cite{1}, in the limit of massless $u,d,s$  quarks $f^2_0$
is related to the first derivative $\chi^{\prime}(0)$  of the correlator of
topological charge densities $Q_5(x)$

\be
\chi(q^2) = i~\int d^4 x\,e^{iqx} \langle 0\mid T\left \{
Q_5(x),~Q_5(0)\right\}\mid 0 \rangle \,,
\label{6}
\ee

\be
Q_5(x) = \frac{\alpha_s}{8\pi}G^n_{\mu \nu} (x)\tilde{G}^n_{\mu\nu}(x),
\label{7}
\ee
where $G^n_{\mu\nu}(x)$  is gluonic field strength, $\tilde{G}^n_{\mu\nu}(x)$ is its
dual, $\tilde{G}^n_{\mu\nu}=(1/2)\varepsilon_{\mu\nu\lambda\sigma} 
G^n_{\lambda\sigma}$:

\be
f^2_0 = 12\chi^{\prime}(0)\,.
\label{8}
\ee
As follows from (\ref{4}),

\be
\chi^{\prime}(0) = (2.3\pm 0.6)\cdot 10^{-3}~GeV^2\,.
\label{9}
\ee
Let us remind the derivation of (\ref{8}). Using (\ref{3})  we can write:

\be
\langle 0 \mid j_{\mu 5} \mid 0 \rangle_A = \lim_{q\to 0} \,i\int d^4
x\,e^{iqx} \langle 0 \mid T \left \{ j_{\nu 5}(x),~ j_{\mu 5}(0) \right \}
\mid 0 \rangle A_{\nu}\equiv
\lim_{q\to 0} P_{\mu\nu}(q) A_{\nu}.
\label{10}
\ee
The general structure of $P_{\mu\nu}(q)$  is

\be
P_{\mu\nu}(q) = -P_L(q^2)\delta_{\mu\nu} +P_T(q^2)(-\delta_{\mu\nu}q^2 +
q_{\mu}q_{\nu}).
\label{11}
\ee
Because of anomaly there are no massless states in the spectrum of singlet
polarization operator $P_{\mu\nu}$  even for massless quarks. $P_{T,L}(q^2)$
also have no kinematical singularities at $q^2=0$. Therefore, the
nonvanishing value $P_{\mu\nu}(0)$  comes entirely from $P_L(q^2)$.
Multiplying $P_{\mu\nu}(q)$ by $q_{\mu}q_{\nu}$ and using the anomaly
condition

\be
\partial_{\mu}j_{\mu 5} (x) = 2N_f Q_5(x)  + 2i \sum_q
m_q\bar{q}(x)\gamma_5 q(x)
\label{12}
\ee
($N_f$ is the number of flavours, $N_f=3$), in the limit of massless quarks
we get:

\be
q_{\mu}q_{\nu}P_{\mu\nu}(q) = -P_L(q^2)q^2 = 36\chi(q^2).
\label{13}
\ee
As is known \cite{3}, $\chi(0)=0$  if there  is at least one massless quark.
(\ref{8})  follows directly from  (\ref{1}), (\ref{10}), (\ref{11}),
(\ref{13}). According to (\ref{1})  we have also:

\be
f^2_0=-(1/3) P_L(0).
\label{14}
\ee

The attempt to determine $f^2_0$  directly by constructing a special QCD sum
rule for this aim was performed in \cite{4}.  However, this attempt failed:
it was found that the operator product expansion (OPE) used in the sum
rule does not converge on the accounted in calculation terms -- the
unaccounted higher order terms of OPE should be of importance. In the
present paper we use the instanton model \cite{5} (as review see \cite{6}) for
calculation of these higher order terms. The idea that instantons give the
main contribution to the longitudinal part $P_L(q^2)$ of the correlator of
singlet axial current $P_{\mu\nu}(q)$  at intermediate $\mid q^2\mid \sim
1~GeV^2$  is not new -- it was suggested as early as in 1979 \cite{7,8}. In
\cite{7}  it was argued that the appearance of $\eta$- and 
$\eta^{\prime}$-mesons as almost pure octet and singlet states in $SU(3)$   flavour
symmetry, i.e. the large mixing of $\bar{u}u + \bar{d}d$ and $\bar{s}s$
in this channel,
cannot be described by perturbative QCD and may be attributed only to
dominating instanton contribution.

The sum rule for direct determination of $f^2_0$ is constructed. At
standard parameters of the instanton model the value of $f^2_0$ found from
the sum rule is in a good agreement with (\ref{4}).

For massless quarks the phenomenological side of the sum rule is saturated
by $\eta^{\prime}$-meson contribution (plus contributions of excited states,
approximated by continuum). The strange quark mass $m_s$ may be also
accounted in the sum rule. In this case the contribution of $\eta$-meson
also comes into play and $\eta - \eta^{\prime}$ mixing angle can be found
from the analysis. In the model of two mixing angles \cite{9} the value of
the largest angle  $\theta_8$ is determined and appears to be in a good
agreement with the values found from the chiral theory and phenomenology
\cite{9,10,11}.

In the framework of the same instanton model the $q^2$-dependence of the
topological charge densities correlator $\chi(Q^2)$ is determined at
space-like $Q^2=-q^2 > 0$. At intermediate $Q^2 \sim 1 GeV^2$ it matches
well with the curve of the $\chi(Q^2)$ behaviour at low $Q^2$, found in
\cite{12} on the basis of (\ref{9}) and contributions of Goldstone bosons
$\pi^0$ and $\eta$.

\section{The sum rule.}

\bigskip

Strictly speaking, the quantity $f_0^2$ given by (\ref{1}) is defined as
nonperturbative part of the induced by the external field $A_{\mu}$ vacuum
expectation value $\langle 0 \mid j_{\mu 5} \mid 0 \rangle_A$ with
perturbative contribution subtracted. The numerical value (\ref{4})
corresponds just to such definition. The reason for this definition is that
in the sum rule for $\Sigma$, from which the value of $f^2_0$ (\ref{4}) was
determined, all perturbative contributions were accounted explicitly and
only nonperturbative part of $\langle 0 \mid j_{\mu 5} \mid 0 \rangle_A$ was
parameterized by unknown constant $f^2_0$. Similarly, $\chi(q^2)$ in
(\ref{6}) and (\ref{8}) has the meaning of nonperturbative part of the
correlator of topological charge densities. Such definition is physically
reasonable, since the perturbative part of (\ref{6}) is badly divergent,
strongly depends on renormalization scheme, and therefore has no physical
meaning. The same statement refers to perturbative contribution to $f^2_0$.
Such separation of perturbative and nonperturbative contributions allows one
to avoid any uncertainties in the sum rule for the physically measurable
quantity $\Sigma$.

The idea of determination of $f^2_0$ or, what is equivalent, proportional
to $f^2_0$ quantity $P_L(0)$ was suggested in \cite{4}. In short, it was the
following.

The imaginary part of $P_L(q^2)$ is represented by contributions of the
lowest resonance -- $\eta^{\prime}$-meson -- and continuum:

\be
Im P_L(q^2) = 3 \pi \tilde{f}^2_{\eta{\prime}} m^2_{\eta{\prime}} \delta(q^2
- m^2 _{\eta^{\prime}})+ \beta(q^2)\theta (q^2 - s_0)\,.
\label{15}
\ee
Here $\tilde{f}_{\eta{\prime}}$ is the coupling of $\eta^{\prime}$-meson
with the singlet axial current

\be
\langle 0 \mid j_{\mu 5} \mid \eta^{\prime} \rangle = i \sqrt{3}
\tilde{f}_{\eta^{\prime}}q_{\mu}
\label{16}
\ee
in the limit of massless $u, d, s$ quarks  ($q_{\mu}$ is $\eta^{\prime}$
momentum). The second term  in the right hand side (rhs) of (15) represents
the contribution of continuum, $s_0$ is the continuum threshold. The
continuum contribution corresponds to gluonic bare loop in the correlator
(\ref{6}) and is equal to

\be
\beta(q^2) = \frac{9\alpha^2_s}{8\pi^3}q^2\,.
\label{17}
\ee
In order to get the nonperturbative part of $Im P_L$ the perturbative part
equal to $\beta(q^2)$ must be subtracted from (15) what gives:

\be
ImP_L(q^2)_{nonp} = 3 \pi \tilde{f}^2_{\eta^{\prime}}m^2_{\eta^{\prime}}
\delta(q^2-m^2_{\eta^{\prime}}) - \beta(q^2)\theta(s_0-q^2)\,.
\label{18}
\ee
As was shown in \cite{4}, the nonperturbative part of $P_L(0)$  is given by

\be
P_L(0)_{nonp}=-3\tilde{f}^2_{\eta^{\prime}} + \frac{1}{\pi}
\int\limits^{s_0}_0\frac{\beta(s)}{s}ds\,.
\label{19}
\ee
Therefore the problem reduces to determination of the coupling constant
$\tilde{f}^2_{\eta^{\prime}}$. This can be done by standard technique of the
QCD sum rule approach. Let us write OPE for $P_L(Q^2)=(36/Q^2)\chi(Q^2)$  at
large $Q^2$. 
We account the instanton contribution in
the instanton liquid approximation. (The instanton contribution was not
accounted in \cite{2}.) The OPE for $P_L(Q^2)$ has the form \cite{8}:

$$-P_L(Q^2)= \frac{9\alpha^2_s}{8\pi^4} Q^2ln {Q^2\over{\mu^2}} +
\frac{9\alpha_s}{4\pi}\frac{1}{Q^2}\langle 0 \mid \frac{\alpha_s}{\pi}G^2
\mid 0 \rangle +$$

$$+ \frac{9\alpha^2_s}{2\pi^2} \frac{1}{Q^4} gf^{abc} \langle 0 \mid
G^a_{\mu \alpha} G^b_{\alpha\beta} G^c_{\beta\mu} \mid 0 \rangle +$$

$$+ \frac{9\alpha^3_s}{2\pi Q^6}f^{abc}f^{ade} \langle 0
\mid G^b_{\mu \nu} G^c_{\alpha\beta} G^d_{\mu\nu} G^e_{\alpha\beta}+10
G^b_{\mu\alpha} G^c_{\alpha\nu} G^d_{\mu\beta} G^e_{\beta\nu}\mid 0 \rangle
+$$

\be
+ 18 Q^2 \int d\rho n(\rho)\rho^4 K^2_2(Q\rho)\,.
\label{20}
\ee
In (\ref{20})  operators up to dimension 8 were accounted. The first term in
the rhs of (\ref{20}) is the bare loop contribution, the last one is the
contribution of instanton \cite{7,8}, $K_2(x)$  is the McDonald function,
$\rho$ is the instanton size and $n(\rho)$  is the instanton density. (In
our normalization anti-instantons are accounted in the coefficient 18 in
front of the last term in (\ref{20}) and $n(\rho)$ represents the instanton
density.)  For $n(\rho)$ we use the Shuryak model \cite{5,6}  of instanton
density:

\be
n(\rho) = n_0\delta(\rho-\rho_c)\,.
\label{21}
\ee
As was  demonstrated by Shuryak and his collaborators \cite{6}, this model
well describes many hadronic correlators in QCD. For numerical values of 
parameters in
(\ref{21})  we choose: $n_0=0.75 \cdot 10^{-3}~GeV^4$,
$\rho_c=1.5~GeV^{-1}$, inside allowed by this model limits.
Particularly, at this $n_0$ the standard value of gluonic condensate

\be
\langle 0 \mid {{\alpha_s}\over {\pi}}G^2 \mid 0 \rangle = 0.012~GeV^4
\label{22}
\ee
may be attributed entirely to instantons. In order to 
estimate 8-dimensional gluonic
condensate we assume the factorization hypothesis -- the saturation by vacuum
intermediate states. Then \cite{8}:

$$
f^{abc}f^{ade}\langle 0 \mid G^b_{\mu\nu}G^c_{\alpha\beta}G^d_{\mu\nu}
G^e_{\alpha\beta} + 10 G^b_{\mu\alpha}
G^c_{\alpha\nu}G^d_{\mu\beta}G^e_{\beta\nu}\mid 0 \rangle=$$
\be
=\frac{15}{16}\langle 0 \mid G^n_{\mu\nu} G^n_{\mu\nu}\mid 0 \rangle^2\,.
\label{23}
\ee
It should be mentioned that the calculation of the same term in the instanton
model would give quite different result:

\be
(2^{11} \cdot 3\pi/7)n_0\rho_c^{-4},
\label{24}
\ee
which is by an order of magnitude larger than (\ref{23}) at accepted model
parameters. This fact is not surprising. Indeed, for the gluonic condensate
with $k$  gluonic fields on dimensional ground we would have:

\be
\langle 0 \mid G^k\mid 0 \rangle \sim \int n(\rho)
\frac{1}{\rho^{2k-4}}d\rho\,,
\label{25}
\ee
and the integral (\ref{25}) diverges at high enough $k$  at any physical
$n(\rho)$. Therefore, one may expect that the instanton model overestimates
the value of 8-dimensional gluonic condendate and accept the estimation
(\ref{23}) based on factorization hypothesis. This estimation is supported
by the analysis of the sum rules for heavy quarkonia with the account of
 8-dimensional operators \cite{13} where the  factorization hypothesis was
used. Much larger values of 8-dimensional gluonic condensates would contradict
the analysis in \cite{13}. Of course, two or, may be, even 3 times larger
values as (\ref{23})  are not excluded, but, luckily, the contribution of
these condensates to the sum rule is small and even its 3 times increasing
does not influence seriously the result. For the 6-dimensional 
gluonic condensate
there is no other independent estimation as given by instanton model
\cite{8}:

\be
\frac{g^3}{12\pi^2}f^{abc}\langle 0 \mid G^a_{\mu\nu}G^b_{\nu\alpha}
G^c_{\alpha\mu} \mid 0 \rangle = \frac{4}{5}\frac{1}{\rho^2_c}
\langle 0 \mid  \frac{\alpha_s}{\pi}G^2\mid 0 \rangle\,.
\label{26}
\ee

The phenomenological representation of $P_L(Q^2)$ follows from (\ref{15}).
Equating these two representation and applying the Borel transformation to
both  sides of this equality we get the sum rule:

$$3\tilde{f}^2_{\eta^{\prime}} m^2_{\eta^{\prime}}
e^{-m^2_{\eta^{\prime}}/M^2}=\frac{9\alpha^2_s}{8\pi^4} M^4E_1\Biggl
(\frac{s_0}{M^2}\Biggr ) + \frac{9\alpha_s}{4\pi}\langle 0 \mid
\frac{\alpha_s}{\pi}G^2 \mid 0\rangle \Biggl ( 1 +
\frac{\varepsilon}{M^2}\Biggr ) +$$

\be
+ \frac{135}{64}\frac{\pi \alpha_s}{M^4} \langle 0 \mid 
\frac{\alpha_s}{\pi}G^2 \mid
0\rangle^2 + 18 n_0\rho^4_c {\cal{B}}_{M^2} Q^2K^2_2(Q\rho_c),
\label{27}
\ee

\be
E_1(x)=1-(1+x)e^{-x},
\label{28}
\ee
where ${\cal{B}}_{M^2}$ means  Borel transform. In (\ref{27}) $\varepsilon$
corresponds to contribution of 6-dimensional gluonic condensate. The
instanton model estimation (\ref{26})  gives $\varepsilon=2.2\,GeV^2$. 
Since we
expect that the instanton model overestimates 6-dimensional gluonic
condensate also, we put $\varepsilon=1\,GeV^2$ and include 
possible uncertainty in
the error. The Borel transformation of McDonald function can be done by
using its asymptotic expansion, what leads to (see \cite{14}):

\be
18n_0\rho^4_c{\cal{B}}_{M^2}Q^2 K^2_2(Q \rho_c)\approx 9n_0 M^3\rho_c^3
\sqrt{\pi}e^{-M^2\rho^2_c}\Biggl ( M^2 \rho_c^2 + \frac{13}{4} +
 \frac{165}{32}\frac{1}{M^2\rho_c^2} \Biggr ).
\label{29}
\ee
In our $M^2$ domain the next terms of the expansion are small. In order to 
verify this fact
numerical calculation was done, using integral representation of
McDonald function.\\
\indent The results of the calculation of $\tilde{f}^2_{\eta^{\prime}}$  
according
to the sum rule (\ref{27}) are plotted in Fig.1. (It was put
$\Lambda_{QCD}=200~MeV$, $\alpha_s$ was calculated in the leading order 
approximation.)
 The standard estimation procedure of the
$M^2$ interval where the sum rule is reliable -- the requirement that
highest order terms of OPE are small -- does not work here, because the
instanton contribution dominates, it comprises about 75--80\% of the total.
So we go to physical arguments. We put the continuum threshold
$s_0=2.5~GeV^2$ close to the position of the second resonance with
$\eta^{\prime}$  quantum numbers, $\eta^{\prime}(1440)$ (probably,
$\eta(1295)$ the belongs to octet), and require that the continuum
contribution to bare loop does not exceed $\sim 50\%$. As a low limit of
$M^2$  interval we choose the $M^2$  value where the $M^2$-dependence
starts to rise steeply. These requirements result in $1.2 \la M^2 \la
1.6~GeV^2$. In this interval $M^2$-dependence is not very strong and we
estimate $\tilde{f}^2_{\eta^{\prime}}\approx(2.4\pm 0.6)\cdot
10^{-2}~GeV^2$. (The error includes 15\% possible variation of $\rho_c$.)
The contribution of the second term in the $rhs$ of (\ref{19})  is
negligibly small. From (\ref{14}) and (\ref{8}) we have finally:

\be
f^2_0=(2.4 \pm 0.6)\cdot 10^{-2}~GeV^2\,,
\label{30}
\ee

\be
\chi^{\prime}(0)=(2.0\pm 0.5)\cdot 10^{-3}~GeV^2\,,
\label{31}
\ee
in a good agreement with (\ref{4}) and $\chi^{\prime}(0)$  value found in
\cite{1}, $\chi^{\prime}(0)=(2.3 \pm 0.6)\cdot 10^{-3}~GeV^2$.

\section{The account of strange quark mass. $\eta^{\prime}-\eta$ mixing
angle.}

Consider the polarization operator $P_L(q^2)$ with the account of the
strange quark mass $m_s$ and determine the coupling constant
$f_{\eta^{\prime}}$ of physical $\eta^{\prime}$. The $u$- and $d$-quark
masses are disregarded as before. Using the definition of $P_{\mu\nu}(q)$
and (\ref{12}) we have:

$$-P_L(q^2)q^2 = q_{\mu}q_{\nu}P_{\mu\nu}(q)=i\int d^4 xe^{iqx}\langle 0 \mid T
 \{ 2N_f Q_5(x),~2N_fQ_5(0)\} + $$

$$+T\{ 2N_fQ_5(x),~ 2im_s\bar{s}(0)\gamma_5 s(0)\} + 
T\{2im_s\bar{s}(x)\gamma_5 s (x),~
 2N_fQ_5(0)\} -$$

\be
-4m_s^2T\{\bar{s}(x)\gamma_5 s(x),~ \bar{s}(0)\gamma_5 s(0)\}\mid 0 \rangle
+4m_s\langle 0 \mid \bar{s}(0)s(0)\mid 0 \rangle\,.
\label{32}
\ee
The last term in (\ref{32})   is caused by the equal-time commutator.
Perform the OPE in the rhs of (\ref{32}). Then in comparison with
(\ref{20}) three  additional terms appear: the equal-time commutator term,
proportional to $m^2_s$ term, corresponding to bare loop of strange quarks,
 and arising from the second and third terms in (\ref{32}) the term,
proportional to quark-gluon condensate \cite{15}:

\be
-g\langle 0 \mid \bar{s}\sigma_{\mu\nu}(\lambda^n/2) G^n_{\mu\nu}
s \mid 0 \rangle = m^2_0 \langle 0\mid \bar{s}s \mid 0 \rangle,
\label{33}
\ee
where $m_0^2=0.8~GeV^2$  was determined in \cite{16}. After Borel
transformation the rhs  of the sum rule is now:

\be
R(M^2)-4m_s\langle 0 \mid \bar{s}s \mid 0 \rangle + \frac{3m^2_s}{2\pi^2}M^2
E_0\Biggl ( \frac{s_0}{M^2}\Biggr ) L^{-8/9}
-6\alpha_s\frac{m_sm^2_0}{\pi M^2}\langle 0 \mid \bar{s}s \mid 0 \rangle,
\label{34}
\ee
where $R(M^2)$ us the rhs of (\ref{27})  and

\be
E_0(x) = 1 - e^{-x},
\label{35}
\ee

\be
L=ln(M^2/\Lambda^2)/ln(\mu^2/\Lambda^2)\,.
\label{36}
\ee
The factor $L^{-8/9}$  accounts quark mass anomalous dimension.

It is useful to consider also the correlator $P_{\mu\nu}(q)$  in the case
when one of the currents is still $j_{\mu 5}(x)$, but the other is the
current of $u$- and $d$-quarks: $\bar{u}\gamma_{\mu}\gamma_5 u+
\bar{d}\gamma_{\mu}\gamma_5 d$. In  this case the rhs of the sum rule is
equal to

\be
\frac{2}{3}R(M^2) - 2\alpha_s \frac{m_sm^2_0}{\pi M^2}\langle 0 \mid
\bar{s}s \mid 0 \rangle\,.
\label{37}
\ee
In the phenomenological -- left hand side (lhs) of the sum rule --
$\eta^{\prime}$- and $\eta$-mesons are contributing and their mixing must be
accounted. We adopt the two mixing angles model \cite{9}, which is based on
the low energy chiral effective theory and describes the experimental data
better \cite{9,10} than the one mixing angle model. In this model the
couplings of $\eta$- and $\eta^{\prime}$-mesons $f_{\eta}$  and
$f_{\eta^{\prime}}$ to octet and singlet axial currents are related to the
couplings of fictitious octet and singlet pseudoscalar states $f_8$ and $f_1$
by:

$$f^8_{\eta} = f_8\,cos\theta_8 ~~~~~~~~~~ f^1_{\eta}=-f_1\,sin\theta_1$$

\be
f^8_{\eta^{\prime}}=f_8\,sin\theta_8 ~~~~~~~~~~~~
f^1_{\eta}=f_1\,cos\theta_1
\label{38}
\ee
The $\eta^{\prime}$- and $\eta$-meson contributions to $P_L(q^2)$ can be
easily calculated in this model. It is convenient to present them separately
for the cases when one of the currents is $\bar{s}\gamma_{\mu}\gamma_5 s$
or $\bar{u}\gamma_{\mu}\gamma_5 u + \bar{d}\gamma_{\mu}\gamma_5 d$ (the
other is always $j_{\mu 5}$). Instead of the lhs of (\ref{27}) we have
now:\\
\indent for $\bar{s}\gamma_{\mu}\gamma_5 s$ current:

$$m^2_{\eta^{\prime}} e^{-m^2_{\eta^{\prime}}/M^2}f^2_1 [~cos^2\theta_1 -
\sqrt{2}(f_8/f_1)sin\theta_8 cos\theta_1~] +$$

\be
+ m^2_{\eta}e^{-m^2_{\eta}/M^2}f^2_1 [~sin^2\theta_1 +
\sqrt{2}(f_8/f_1)sin\theta_1 cos\theta_8~];
\label{39}
\ee
\indent for $\bar{u}\gamma_{\mu}\gamma_5 u + \bar{d}\gamma_{\mu} \gamma_5 d$
current:

$$m^2_{\eta^{\prime}}e^{-m^2_{\eta^{\prime}}/M^2} f^2_1 [~2cos^2\theta_1 +
\sqrt{2}(f_8/f_1)sin\theta_8\,cos\theta_1~] +$$

\be
+m^2_{\eta}e^{-m^2_{\eta}/M^2} f^2_1~[~2sin^2\theta_1 - \sqrt{2}(f_8/f_1)
cos\theta_8\,sin\theta_1~]\,.
\label{40}
\ee
Taking the sum of (\ref{39}), (\ref{40}), putting $\theta_1=\theta_8=0$ and
equating it to (\ref{34}) at $m_s=0$,  we get the previous result with
$f_1=f_0$. At nonzero $m_s$ the mixing angles must be accounted and for the
sum of (\ref{39}), (\ref{40}) we get:

$$3m^2_{\eta^{\prime}}f^2_1 ~e^{-m^2_{\eta^{\prime}}/M^2}\Biggl [
cos^2\theta_1 + \frac{m^2_{\eta}}{m^2_{\eta^{\prime}}}
e^{(m^2_{\eta^{\prime}}-m^2_\eta)/M^2} sin^2\theta_1 \Biggr ] =$$

\be
=R(M^2) - 4m_s \langle 0\mid \bar{s}s\mid 0 \rangle + \frac{3m^2_s}{2\pi^2}M^2
E_0\Biggl ( \frac{s_0}{M^2}\Biggr ) L^{-8/9} - 6 \alpha_s \frac{m_s m
^2_0}{\pi M^2}\langle 0 \mid \bar{s}s \mid 0 \rangle\,.
\label{41}
\ee
Take now the difference of (\ref{39})  and one half of (\ref{40}). The
corresponding sum   rule is:

$$-\frac{3}{\sqrt{2}} m^2_{\eta^{\prime}}e^{-m^2_{\eta_{\prime}}/M^2}
f_1f_8\Biggl [ sin \theta_8\, cos\theta_1 -
\frac{m^2_{\eta}}{m^2_{\eta^{\prime}}} e^{(m^2_{\eta^{\prime}}-m^2_\eta)/M^2}
cos\theta_8\,sin\theta_1~\Biggr ] =$$

\be
=-4m_s\langle 0 \mid \bar{s}s \mid 0 \rangle + \frac{3m^2_s}{2\pi^2}M^2
E_0\Biggl (\frac{s_0}{M^2} \Biggr ) L^{-8/9} -3\alpha_s \frac{m _s
m^2_0}{\pi M^2}\langle 0 \mid \bar{s}s \mid 0 \rangle \,.
\label{42}
\ee
The theoretical value of $\theta_1$ found in \cite{9,10,11} is small:
$\theta_1=-(2.7^0-4^0)$. (The phenomenological value \cite{10,11} is
slightly higher: $\theta_1=-9.2^0$). Therefore, with a good accuracy we can
put $\theta_1\approx 0$ in (\ref{41}),(\ref{42}). Then (\ref{41})
determines $f^2_1\approx f^2_{\eta^{\prime}}$. $M^2$-dependence of $f^2_1$
is presented on Fig.1. (The numerical values $\langle 0\mid \bar{s}s \mid
0\rangle=-1.11 \cdot 10^{-2}~GeV^3$ and $m_s(1~GeV)=150~MeV$ were used). From 
the curve in Fig.1 the estimation follows:

\be
f^2_{\eta^{\prime}}=(3.2 \pm 0.6)\cdot 10^{-2}~GeV^2,
~~~f_{\eta^{\prime}}=178 \pm 17~MeV.
\label{43}
\ee
The ratio of (\ref{42}) to (\ref{41})  gives the value of mixing angle
$\theta_8$. In the approximation $\theta_1\approx 0$ and at $f_8/f_1=1.12$
\cite{9,10,11} it is equal:

\be
\theta_8=-(17.0 \pm 5.0)^0.
\label{44}
\ee
The account of $\theta_1=-2.7^0$  changes $\theta_8$ to

\be
\theta_8=-(18.8\pm 5.0)^0.
\label{45}
\ee
The values of $f_{\eta^{\prime}}$  and $\theta_8$  mixing angle are in an
agreement with ones found in \cite{9,10,11}  from the low energy
effective theory or phenomenology.  They are correspondingly:

$$\mbox{theory:} ~~~~~~~ f_{\eta^{\prime}} = 151 ~MeV; ~~~~~\theta_8=-21^0\,.$$

\be
\mbox{phenom.:} ~~~~~~ f_{\eta^{\prime}} = 153 ~MeV; ~~~~~\theta_8=-21^0\,.
\label{46}
\ee

\section{$Q^2$-dependence of $\chi(Q^2)$.}

\bigskip
Using OPE for $P_L(Q^2)$ --  (\ref{20}) and (\ref{13}), the
$Q^2$-dependence of $\chi(Q^2)$ at high $Q^2$  can be found. Since, by
definition of $\chi(Q^2)$, the perturbative  part should be
subtracted, the first term in the rhs of (\ref{20}) is omitted
and we have:

$$\chi(Q^2) = -\frac{\alpha_s}{16\pi} \langle 0 \mid \frac{\alpha_s}
{\pi} G^2 \mid 0 \rangle \Biggl (1+\frac{\varepsilon}{Q^2} \Biggr ) -
\frac{15}{128} \pi \alpha_s\frac{1}{Q^4}
\langle 0\mid \frac{\alpha_s}{\pi} G^2 \mid 0\rangle^2-$$

\be
- \frac{1}{2} n_0Q^4 \rho^4_c K^2_2(Q\rho_c),
\label{47}
\ee
where $\varepsilon$ parameterizes the 6-dimensional gluonic condensate
contribution
(see (\ref{27})) and (\ref{23}) was used. $\chi(Q^2)$ (\ref{47}) is plotted
in Fig.2. It is instructive to compare $\chi(Q^2)$ at large $Q^2$
with $\chi(Q^2)$
at low $Q^2$ found in \cite{12}:

\be
\chi(Q^2)=\chi(0) - \chi^{\prime}(0)Q^2 - \frac{1}{8} f^2_\pi Q^2 \Biggl
[ \Biggl ( \frac{m_u-m_d}{m_u+m_d}\Biggl )^2
\frac{m^2_{\pi}}{Q^2+m^2_{\pi}} + \frac{1}{3}
\frac{m^2_{\eta}}{Q^2+m^2_{\eta}} \Biggr ],
\label{48}
\ee
where

\be
\chi(0) = \frac{m_um_d}{m_u+m_d} \langle 0 \mid \bar{u} u \mid 0 \rangle
\label{49}
\ee
and the last term represents the contributions of $\pi^0$- and
$\eta$-mesons. The curve of the low $Q^2$ behaviour of $\chi (Q^2)$
(\ref{48}) is also plotted in Fig.2, for $\chi^{\prime} (0)$ it was
chosen the value found in \cite{1}: $\chi^{\prime} (0) = 2.3\cdot 10^{-3}
~GeV^2$. As is seen from Fig.2, both curves matches rather well in the
domain $Q^2 \approx 0.4 - 1\, GeV^2$.

\section{Discussion. Comparison with results of other works.}

\bigskip

As was mentioned in the Introduction, instantons are the most plausible
QCD objects for description of physical $\eta^\prime$ channels, or, what is
equivalent, of the longitudinal part of singlet axial vector current
correlator. The calculations of the pseudoscalar current $j_5=
\bar{q} \gamma_5 q$ correlator with $\eta^{\prime}$ quark content
performed by Shuryak and Verbaarschot \cite{17} and Sch\"afer \cite{18}
in the framework of various instanton models demonstrated that this
correlator is described in an agreement with its phenomenological
coordinate dependence up to distances $x \sim 0.3$ fermi (and in some
of them even up to larger ones). Therefore, beforehand, we could expect
that the instanton model is suitable for consideration of the problem
in view. However, in this paper we used the simplest version of the
instanton model -- the instanton liquid approximation with instanton
density given by the Shuryak model (\ref{21}). For this reason the
accuracy of our results is limited.

Since the main contribution (about 80 \%) to the sum rules from which
$f^2_0$ and $f^2_{\eta{\prime}}$ were found comes from the instanton
term, the main uncertainty is caused by the instanton parameters $n_0$
and $\rho_c$. These parameters were taken from the best fit to various
hadronic correlators, as well as some other QCD objects, like gluonic
condensate performed by Shuryak and his collaborators \cite{6}.
Possible uncertainties are included into the errors. The errors in the
determination of the mixing angle $\theta_8$ are smaller, because
instantons do not contribute to the rhs of (42). If for $f_1$ and $f_8$
their phenomenological values will be taken: $f_1 = 1.28 f_{\pi}, f_8 =
1.15 f_{\pi}$ \cite{9,10,11} instead of $f_1 = f_{\eta^{\prime}}$ found
from the sum rule, 
then we would have from (\ref{42}) (at $\theta_1 = -2.7^0$):

\be
\theta_8 = -(26.5 \pm 3.5)^0\,.
\label{50}
\ee
The obtained above value of $f_{\eta^{\prime}}$ (\ref{43}) is a bit
higher than the low energy data (\ref{46}), but, taking in mind the
uncertainties, one may consider the agreement as satisfactory. \\
\indent The
$\alpha_s$-corrections were not accounted in our calculation. They are
essential in the case of the first term in the rhs of (\ref{27})
\cite{19}, but this term contributes only 5\% to the total result. The
$\alpha_s$-corrections to gluonic condensate contributions are masked
by uncertainties in higher order gluonic condensates. Among the terms
proportional to $m_s$ and $m^2_s$, the $\alpha_s$-corrections appear
only to the two last terms in (\ref{34}), not to the main,
proportional to $m_s$, term $\;-4 m_s \langle 0 \mid \bar{s}s \mid 0
\rangle$. They do not much influence the value of the mixing angle.
There are also instanton corrections to the term $\sim m^2_s$ of the
same order of magnitude. The account of all these corrections to the
$\sim m^2_s$ terms is the subject of further study. We believe that
after their account the value of the mixing angle will be still in the
limits of accepted above errors.\\
\indent The slopes in $M^2$ of the left and right
hand sides of (\ref{41}) are different -- positive in the lhs and
negative in the rhs. For this reason it is impossible to determine
$m^2_{\eta^{\prime}}$  by differentiation of (\ref{41}) in $M^2$ as it
was done sometimes in the QCD sum rule approach: whereas the sum rule
is satisfactory, its derivative is not.

In their recent paper Narison, Shore and Veneziano \cite{19} studied
the problem of determination of $\chi^{\prime}(0)$ or
$f_{\eta{\prime}}$ in the framework of the standard QCD sum rule
approach with no account of instantons. Their result for
$\chi^{\prime}(0)$ or $f^2_{\eta^{\prime}}$ is essentially -- 3-4 times --
smaller than ours. The principal difference from the presented here
calculations (besides instanton contribution) is that the authors of
\cite{19} have chosen much larger values of the continuum threshold
$s_0$ and of the effective Borel parameters: $s_0 = 6\, GeV^2$ and $M^2
\sim 3-4\, GeV^2$. Therefore, they accepted the model of hadronic
spectrum in $J^{PC} = 0^{-+}$ flavour singlet channel with a gap between
$\eta^{\prime}$-meson mass and $2.5\,GeV$. However, there are at least 3
resonances with $\eta^{\prime}$ quantum numbers between $\eta^{\prime}$
mass and 2.5 GeV. That is why we think that such a model is not
acceptable physically. The other drawback of the sum rule used in
\cite{19} (eq.(D.11) of \cite{19},  which is similar to (\ref{41}), but
without $\eta - \eta^{\prime}$ mixing) is that the main contribution to
the sum rule comes from the terms proportional to $m_s$ and $m^2_s$.
These terms comprise 60\% of the final answer. This means that SU(3)
flavour symmetry is badly violated in the sum rule \cite{19} in
contradiction with experiment. Moreover, if we introduce $\eta -
\eta^{\prime}$ mixing (what was not done in \cite{19}), we may
calculate the $\eta - \eta^{\prime}$ mixing angle, representing the
phenomenological side of the sum rule by (\ref{39}),(\ref{40}).
 Then the result for the mixing angle $\theta_8$ is: $\theta_8
\approx 45^0$, i.e. $\eta^{\prime}$ is not mainly flavour singlet and
$\eta$ is not mainly octet -- in evident contradiction with experiment.

In conclusion, we have shown that the instanton model even in its
simplest version describes reasonably well the properties of the
topological density correlator -- $\chi^{\prime}(0)$ and $\chi(Q^2)$ at
large $Q^2$, the values of the $\eta^{\prime}$ coupling constant and 
the $\eta^{\prime}- \eta$ mixing angle.

This work was supported in part by RFBR grant 97-02-16131.



\newpage
\bf
$$\centerline{\hbox{\Large{Figures.}}}$$
\rm
\unitlength=1.00mm
\special{em:linewidth 0.4pt}
\linethickness{0.4pt}
\begin{picture}(120.00,130.00)
\put(50.00,80.00){\vector(1,0){70.00}}
\put(128.00,75.00){\makebox(0,0)[cc]{\small{$M^2,GeV^2$}}}
\put(116.00,96.00){\makebox(0,0)[cc]{\small{$\tilde{f}^2_{\eta^\prime}$}}}
\put(116.00,103.00){\makebox(0,0)[cc]{\small{$f^2_{\eta^\prime}$}}}
\bezier{276}(60.00,115.00)(74.00,105.00)(110.00,99.00)
\bezier{276}(60.00,122.00)(74.00,112.00)(110.00,105.00)
\put(50.00,80.00){\vector(0,1){50.00}}
\put(62.00,78.00){\line(0,1){4.00}}
\put(74.00,82.00){\line(0,-1){4.00}}
\put(86.00,78.00){\line(0,1){4.00}}
\put(98.00,82.00){\line(0,-1){4.00}}
\put(110.00,78.00){\line(0,1){4.00}}
\put(48.00,90.00){\line(1,0){4.00}}
\put(52.00,100.00){\line(-1,0){4.00}}
\put(52.00,120.00){\line(-1,0){4.00}}
\put(48.00,110.00){\line(1,0){4.00}}
\put(74.00,75.00){\makebox(0,0)[cc]{\small{1.2}}}
\put(86.00,75.00){\makebox(0,0)[cc]{\small{1.3}}}
\put(98.00,75.00){\makebox(0,0)[cc]{\small{1.4}}}
\put(110.00,75.00){\makebox(0,0)[cc]{\small{1.5}}}
\put(62.00,75.00){\makebox(0,0)[cc]{\small{1.1}}}
\put(43.00,90.00){\makebox(0,0)[cc]{\small{0.01}}}
\put(43.00,100.00){\makebox(0,0)[cc]{\small{0.02}}}
\put(43.00,110.00){\makebox(0,0)[cc]{\small{0.03}}}
\put(43.00,120.00){\makebox(0,0)[cc]{\small{0.04}}}
\put(41.00,129.00){\makebox(0,0)[cc]{\small{$GeV^2$}}}
\put(80.00,65,00){\makebox(0,0)[cc]{Figure\,1.}}
\put(30.00,-20.00){\vector(1,0){110.00}}
\put(140.00,-25.00){\makebox(0,0)[cc]{\small{$Q^2,GeV^2$}}}

\bezier{276}(30.00,-20.00)(35.00,15.00)(42.00,25.00)
\bezier{276}(70.00,25.00)(100.00,0.00)(126.00,-3.00)
\bezier{176}(45.00,29.00)(47.00,31.50)(50.00,32.80)
\bezier{176}(62.00,31.20)(64.00,29.60)(66.00,28.00)
\bezier{176}(53.00,33.80)(56.00,34.80)(58.00,33.40)

\put(30.00,-20.00){\vector(0,1){60.00}}
\put(38.00,-18.00){\line(0,-1){4.00}}
\put(46.00,-18.00){\line(0,-1){4.00}}
\put(54.00,-18.00){\line(0,-1){4.00}}
\put(62.00,-18.00){\line(0,-1){4.00}}
\put(70.00,-18.00){\line(0,-1){4.00}}
\put(78.00,-18.00){\line(0,-1){4.00}}
\put(86.00,-18.00){\line(0,-1){4.00}}
\put(94.00,-18.00){\line(0,-1){4.00}}
\put(102.00,-18.00){\line(0,-1){4.00}}
\put(110.00,-18.00){\line(0,-1){4.00}}
\put(118.00,-18.00){\line(0,-1){4.00}}
\put(126.00,-18.00){\line(0,-1){4.00}}

\put(32.00,-10.00){\line(-1,0){4.00}}
\put(32.00,00.00){\line(-1,0){4.00}}
\put(32.00,10.00){\line(-1,0){4.00}}
\put(32.00,20.00){\line(-1,0){4.00}}
\put(32.00,30.00){\line(-1,0){4.00}}

\put(38.00,-25.00){\makebox(0,0)[cc]{\small{0.2}}}
\put(46.00,-25.00){\makebox(0,0)[cc]{\small{0.4}}}
\put(54.00,-25.00){\makebox(0,0)[cc]{\small{0.6}}}
\put(62.00,-25.00){\makebox(0,0)[cc]{\small{0.8}}}
\put(70.00,-25.00){\makebox(0,0)[cc]{\small{1.0}}}
\put(78.00,-25.00){\makebox(0,0)[cc]{\small{1.2}}}
\put(86.00,-25.00){\makebox(0,0)[cc]{\small{1.4}}}
\put(94.00,-25.00){\makebox(0,0)[cc]{\small{1.6}}}
\put(102.00,-25.00){\makebox(0,0)[cc]{\small{1.8}}}
\put(110.00,-25.00){\makebox(0,0)[cc]{\small{2.0}}}
\put(118.00,-25.00){\makebox(0,0)[cc]{\small{2.2}}}
\put(126.00,-25.00){\makebox(0,0)[cc]{\small{2.4}}}

\put(25.00,-10.00){\makebox(0,0)[cc]{\small{0.2}}}
\put(25.00,00.00){\makebox(0,0)[cc]{\small{0.4}}}
\put(25.00,10.00){\makebox(0,0)[cc]{\small{0.6}}}
\put(25.00,20.00){\makebox(0,0)[cc]{\small{0.8}}}
\put(25.00,30.00){\makebox(0,0)[cc]{\small{1.0}}}

\put(45.00,41.00){\makebox(0,0)[cc]{\small{$-\chi\cdot 10^{3}\,GeV^2$}}}
\put(80.00,-35,00){\makebox(0,0)[cc]{Figure\,2.}}
\end{picture}

\newpage
\bf
$$\centerline{\hbox{\Large{Figure captions.}}}$$
\rm

$\;\;$\\
\indent Figure 1.
Functions $\tilde{f}^2_{\eta^\prime}(M^2),\,f^2_{\eta^\prime}(M^2)$\, 
determined by equations (27), (41) correspondingly. In (41) $\theta_1=0$.
\\ \indent Figure 2.
Function $\chi(Q^2)$ at low and large $Q^2$ -- solid lines 
(see equations (48) and (47) 
correspondingly). The dashed line represents matching curve drawn by hand. 

\end{document}